\documentclass[journal]{IEEEtran}
\usepackage{amsmath,amsfonts}
\usepackage{algpseudocode}
\usepackage{algorithm}
\usepackage{array}
\usepackage[caption=false]{subfig}
\usepackage{textcomp}
\usepackage{stfloats}
\usepackage{url}
\usepackage{verbatim}
\usepackage{graphicx}
\usepackage{xcolor}
\usepackage[hidelinks]{hyperref}
\usepackage{multirow}
\usepackage{tabularx}
\usepackage{amsthm}
\usepackage{hyperref}

\usepackage{cleveref}
\crefname{equation}{}{}
\Crefname{equation}{Equation}{Equations}
\crefname{figure}{Fig.}{Figs.}
\crefname{table}{Table}{Tables}
\crefname{section}{Section}{Sections}
\crefname{algorithm}{Algorithm}{Algorithms}

\newtheorem{lemma}{Lemma}

\allowdisplaybreaks

\title{Data Driven Approach towards More Efficient Newton-Raphson Power Flow Calculation for Distribution Grids
}

\author{Shengyuan Yan\textsuperscript{1}, Farzad Vazinram\textsuperscript{2}, Zeynab Kaseb\textsuperscript{3}, Lindsay Spoor\textsuperscript{4}, Jochen Stiasny\textsuperscript{3}, Bet\"ul Mamudi\textsuperscript{3}, Amirhossein Heydarian Ardakani\textsuperscript{5}, Ugochukwu Orji\textsuperscript{6}, Pedro P. Vergara\textsuperscript{3}, Yu Xiang\textsuperscript{7,8}, Jerry Guo\textsuperscript{7}
\thanks{

\textsuperscript{1}{Mathematics and Computer Science, Eindhoven University of Technology, P.O. Box 513, 5600 MB, Eindhoven, The Netherlands}

\textsuperscript{2}{Computer Science, University of Twente, P.O. Box 217, 7500 AE, Enschede, The Netherlands}

\textsuperscript{3}{Electrical Sustainable Energy, Delft University of Technology, P.O. Box 5031, 2600 GA Delft, The Netherlands}

\textsuperscript{4}{Leiden Institute of Advanced Computer Science, Leiden University, P.O. Box 9512, 2300 RA Leiden, The Netherlands}

\textsuperscript{5}{Engineering Technology, University of Twente, P.O. Box 217, 7500 AE, Enschede, The Netherlands}

\textsuperscript{6}{Jheronimus Academy of Data Science, Tilburg University, P.O. Box 90153, 5037 AB, Tilburg, The Netherlands}

\textsuperscript{7}{Alliander N.V., P.O. Box 50, 6920 AB, Arnhem, The Netherlands}

\textsuperscript{8}{Electrical Engineering, Eindhoven University of Technology, P.O. Box 513, 5600 MB, Eindhoven, The Netherlands}
}
}

\begin{document}

\maketitle

\begin{abstract}
Power flow (PF) calculations are fundamental to power system analysis to ensure stable and reliable grid operation. The Newton-Raphson (NR) method is commonly used for PF analysis due to its rapid convergence when initialized properly. However, as power grids operate closer to their capacity limits, ill-conditioned cases and convergence issues pose significant challenges. This work, therefore, addresses these challenges by proposing strategies to improve NR initialization, hence minimizing iterations and avoiding divergence. We explore three approaches: (i) an analytical method that estimates the basin of attraction using mathematical bounds on voltages, (ii) Two data-driven models leveraging supervised learning or physics-informed neural networks (PINNs) to predict optimal initial guesses, and (iii) a reinforcement learning (RL) approach that incrementally adjusts voltages to accelerate convergence. These methods are tested on benchmark systems. This research is particularly relevant for modern power systems, where high penetration of renewables and decentralized generation require robust and scalable PF solutions. In experiments, all three proposed methods demonstrate a strong ability to provide an initial guess for Newton-Raphson method to converge with fewer steps. The findings provide a pathway for more efficient real-time grid operations, which, in turn, support the transition toward smarter and more resilient electricity networks.    
\end{abstract}


\section{Introduction}
Electricity, discovered in the nineteenth century, is one of humanity’s most important achievements. Since the initial installation of electricity systems, the distribution grid has evolved significantly in order to adapt to the changing demands of society and advances in technology \cite{moreno2021,majeed2021}:
\begin{itemize}
    \item Rise of electrification 
    \item Digitalized society
    \item Switch from centralized to decentralized generation
\end{itemize}

A critical component of grid operations and control is power flow calculations. These are essential to determine the steady-state operation of the grid and maintain system stability. To solve power flow calculations, the Newton-Raphson approach is often applied, as it is able to find a good solution (converge) with only a few iterations if the initial value is selected properly \cite{okhuegbe2024}.

However, as the grid operates closer to its maximum capacity, particularly during peak demand or high renewable generation, these calculations and the search for a good initial value become increasingly difficult. Ill-conditioned cases at maximum capacity, convergence issues, and the sheer computational complexity of solving power flow equations at scale make efficient and accurate power flow solutions vital to ensure a resilient and reliable electricity system \cite{duque2024}.

This work aims to avoid starting Newton-Raphson with ill-conditioned initial values. We introduce the relevant background in \cref{sec:background} and consider three approaches in \cref{sec:approaches}: 1) Bounding the initial condition based on analytical conditions (\cref{subsec:analytical_approach}); 2) Learning approximate solutions that can be used as improved initial values (\cref{subsec:learning_approximations}); 3) Performing initial update steps with an RL agent (\cref{subsec:RL_based_updates}). \Cref{sec:Futurework} provides future steps to deepen the research of each proposed method. \Cref{sec:conclusion} discusses the results and concludes this research.

The contributions of our work can be summarized in three folds:
\begin{enumerate}
    \item We employed an analytical method to estimate the basin of attraction for Newton-Raphson to limit the input space for the initial guesses. The analytical method does not need any training data and can work with electrical networks of any scale and complexity with some empirical expert experience about setting the proper contraction parameters.
    \item We proposed a solution prediction pipeline using both supervised deep learning and unsupervised physics-informed neural network (PINN) models. The supervised learning model can learn from labeled data and predict rough solutions of PF equations, which are later refined by Newton-Raphson method. The PINN model can learn on unlabeled data and generate physically plausible solutions for PF equations as initial guess for Newton-Raphson method. In some cases, initializing the Newton-Raphson method using a flat start fails to converge. In those instances, the PINN method can provide initial guesses for test samples to converge in a limited number of steps.
    \item We proposed a reinforcement learning (RL) method to move a given initial guess to the area in the initial guesses input space where Newton-Raphson method can easily converge. Experiments show that even when given an initial guess that is very hard for Newton-Raphson method to converge, our RL method can move it to easy-converge area in at most 10 RL agent steps. Using the initial guesses updated by the RL agent, Newton-Raphson method can converge within 10 iterations.
\end{enumerate}

\section{Problem definition}\label{sec:background}

\subsection{Power Flow}

The power flow equations for a network with $N$ buses can be compactly written as
\begin{align}
\bar{S} = \bar{V} \cdot \left(\bar{Y} \bar{V} \right)^*\label{eq:power_flow}
\end{align}
where $\bar{S}$ represents the complex power injections $\bar{S} = P + jQ$ at the buses. They are composed of the active power $P \in \mathbb{R}^N$ and the reactive power $Q\in \mathbb{R}^N$. The imaginary unit is represented by $j$ to avoid confusion with the current $i$. $(\cdot)^*$ represents the complex conjugate. The complex voltage phasor $\bar{V}=V e^{j\theta}$ is expressed by the voltage magnitude $V \in \mathbb{R}^N$ and the voltage angle $\theta \in \mathbb{R}^N$ at each bus. The bus admittance matrix $\bar{Y} = G + jB$ can be separated into its real and imaginary parts, $G \in \mathbb{R}^{N\times N}$ and $B \in \mathbb{R}^{N\times N}$, and it describes the network that transmits the power \cite{milanoPowerSystemModelling2010}.

\subsection{Solving root-finding problems with Newton-Raphson}
\begin{figure}[t!]
    \centering
    \includegraphics[width=0.8\linewidth]{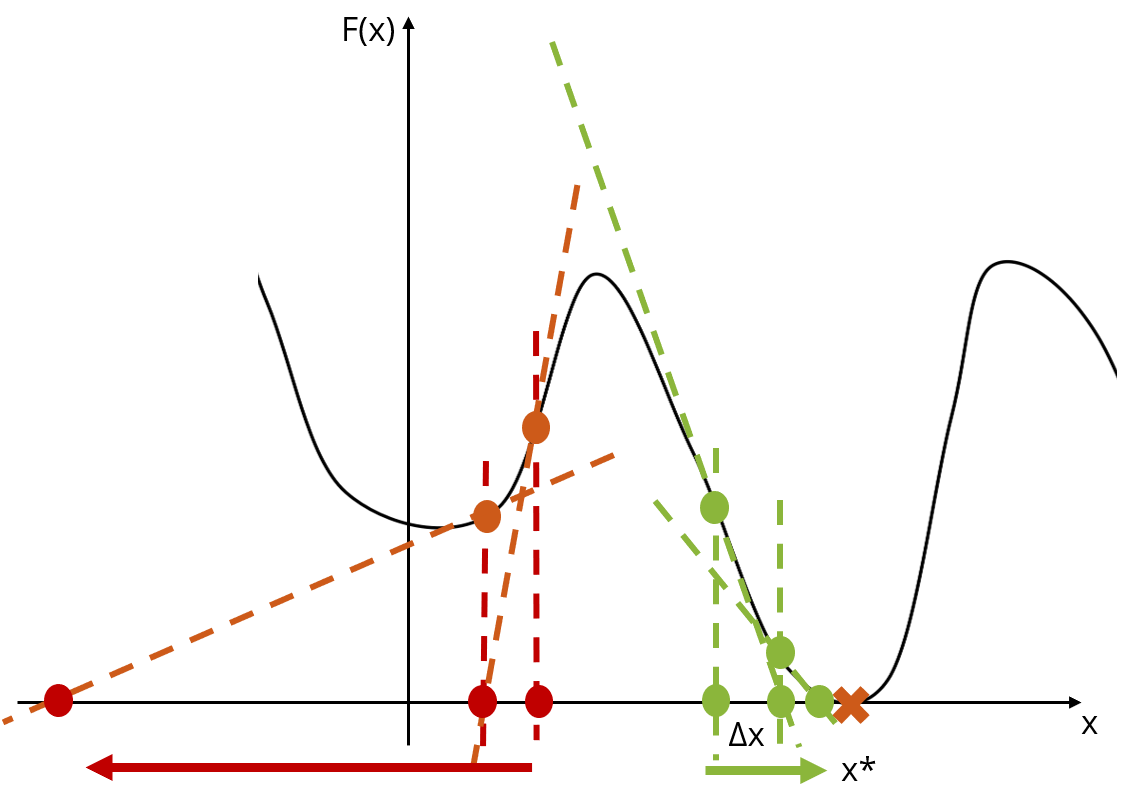}
    \caption{Newton-Raphson method and the contrast between the initial guesses of easy-convergence case and ill-conditioned cases. The green lines and dots represent an example of the updating iterations of easy-convergence initial guess; the red is an example of the updating iterations of ill-conditioned initial guess. The cross marks the true solution, which is the target of Newton-Raphson method.}
    \label{fig_NR_method}
\end{figure}

We will express the power flow problem \cref{eq:power_flow} as a root-finding problem. In such a problem, $x$ represents the unknown variables and $F(x)$ the residual of the governing equations that should equal $0$. To solve such a problem, we apply an iterative procedure as follows:
\begin{align}
    x^{(k+1)} = x^{(k)} + \Delta x,
\end{align}
which startes from an initial value $x^{(0)}$. The Newton-Raphson scheme defines the update step $\Delta x$ as:
\begin{align}
    \Delta x = - \left( \left.\frac{\partial F\left(x\right)}{\partial x}\right|_{x^{(k)}} \right)^{-1} F\left(x^{(k)}\right),
\end{align}
where $\partial F/\partial x$ represents the Jacobian of the residual with respect to the unknown variables. The iterative procedure is considered converged when a tolerance criterion $||F\left(x^{(k)}\right)|| \leq \epsilon$ is met. The Newton-Raphson method iterative updating process is illustrated in Fig. \ref{fig_NR_method}. We consider initial values $x^{(0)}$ that require a large number of iterations to converge, here 10, as \textit{ill-conditioned cases}. Examples of the iterative updating process of easy-convergence and ill-conditioned cases are also present in Fig. \ref{fig_NR_method}.

\subsection{Case Studies}

\begin{figure}[t!]
    \centering
    \includegraphics[width=1.0\linewidth]{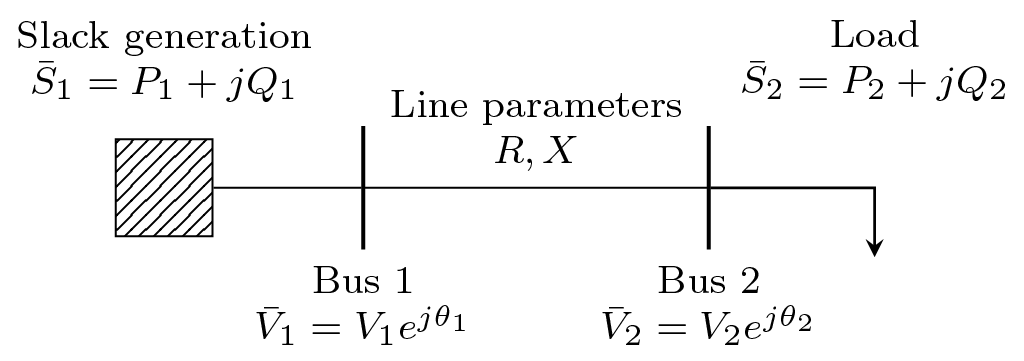}
    \caption{Schematic of the two-bus system.}
    \label{fig:two_bus_system}
\end{figure}

To test our approaches, we apply the methods to a two-bus system shown in \cref{fig:two_bus_system}. The system connects the generation $\Bar{S}_1$ at bus 1 with a load $\Bar{S}_2$ at bus 2 via a line with parameters $R$ and $X$. Bus 1 serves as the slack bus where the voltage is fixed at $\bar{V}_1 = 1.0e^{j0}$ and the generation $\Bar{S}_1$ will be set to match the load. The unknown variables $x$ in this setup are the voltage magnitude $V_2$ and the voltage angle $\theta_2$ at bus 2.
Variations in the case can be generated by altering the load $S_2 = P_2 + jQ_2$ or the line parameters $R$ and $X$.

\begin{figure}[t!]
    \centering
    \includegraphics[width=1.0\linewidth]{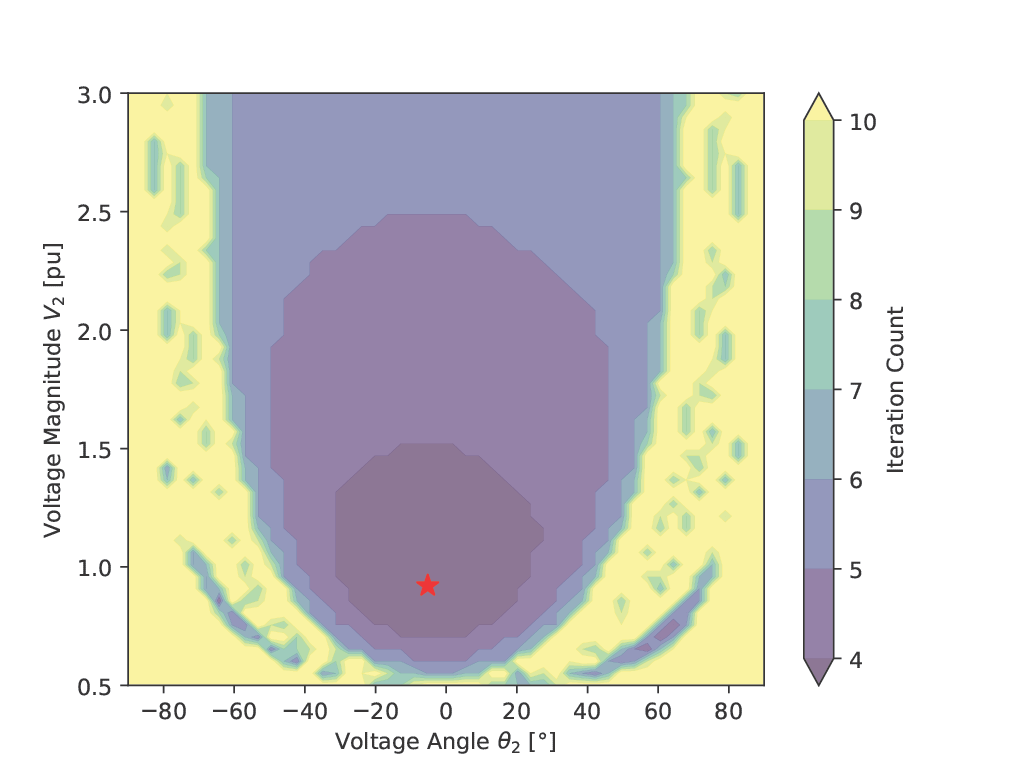}
    \caption{Convergence map for a two-bus system. The axes show the initial values for the voltage angle $\theta_2$ and the voltage magnitude $V_2$ and the colors indicate the number of Newton-Raphson iterations until the desired solution (marked by the star) is reached up to tolerance $\epsilon$.}
    \label{fig:newton_raphson_iterations}
\end{figure}

We illustrate the convergence behavior of Newton-Raphson for this case in \cref{fig:newton_raphson_iterations}. The axes show the initial values for $V_2$ and $\theta_2$, and the color indicates how many iterations are required until convergence. The star marks the solution where the power flow is satisfied. Around this solution, a \textit{basin of attraction} can be observed where Newton-Raphson converges quickly and reliably. 
We also consider a seven-bus system described as ``Simple Example Network'' in the documentation of pandapower \cite{thurner2018pandapower} as a larger test case.

\section{Investigated Approaches}\label{sec:approaches}

\subsection{Analytical Method}\label{subsec:analytical_approach}

One method to cope with the convergence speed of power flow calculation is estimating the basin of attraction. With this method, the convergence region of the power flow algorithm would be defined by estimating the maximum and minimum grid voltage. There are two proved theorems as follows:

\textbf{Theorem 1} \cite{Garces-Ruiz2024-qx,Advanced_calculus_undated-ai}: $\mathcal{B}$ is defined as follows and is a closed ball in a complete metric space X,
\begin{equation}
    \mathcal{B} = \{ V \in \mathbb{C}^n : \| V - \bar{V} \|_\infty \leq r \}
    \label{eq:B}
\end{equation}
and T : $\mathcal{B}$ $\rightarrow$ $\chi$ is a contraction that shifts the center of B by a distance no greater than (1 $-$ $\gamma$)r, where $\gamma$ is the contraction constant, V is voltage magnitude of nodes (except slack bus), and r is the radius of $\mathcal{B}$. Then T has a unique fixed point in $\mathcal{B}$.

\textbf{Theorem 2} \cite{Garces-Ruiz2024-qx}: $\mathcal{B}$ is defined as equation \ref{eq:B} with radius r and center in $\bar{V} = V1$, with $V > 0$. The maximum and minimum roots of the following polynomial will give the $r_{min}$ and $r_{max}$ of the basin:
\begin{equation}
    r^3 - (2\nu + \alpha_2)r^2 + (\nu^2 + 2\alpha_2 \nu - \alpha_1)r - \alpha_2 \nu^2 = 0
    \label{eq:placeholder}
\end{equation}
where,
\begin{equation}
   \alpha_1 = \left\| Z_{\mathcal{N}} s^*_{\mathcal{N}} \right\|_{\infty}
   \label{eq:alpha_1}
\end{equation}
\begin{equation}
\alpha_2 = \left\| \overline{V}_{\mathcal{N}} - Z_{\mathcal{N}} \left( \left( \frac{s_{\mathcal{N}}}{\overline{V}} \right)^* - Y_{\mathcal{M}} v_S \right) \right\|_{\infty}
\label{eq:alpha_2}
\end{equation}

\begin{figure}[t!]
    \centering
    \includegraphics[width=0.7\linewidth]{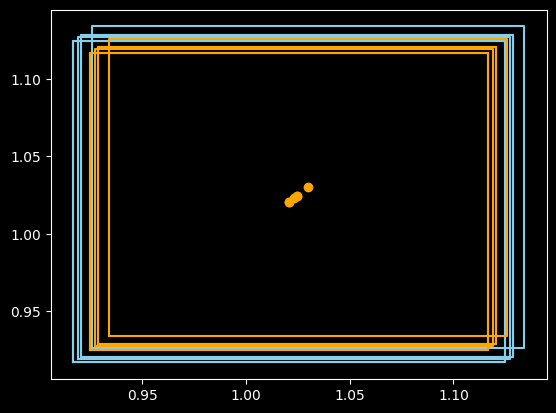}
    \caption{Estimating of the basin of attraction for voltage, including the center and maximum (blue squares) and minimum (orange squares) radius of the region for a system with 7 buses. This method can be applied to grids of any size and configuration, regardless of the number of buses.}
    \label{fig:Basin_of_attraction}
\end{figure}

The $V$ value, as the center of the region, can be selected based on nominal voltage, slack bus voltage, or the voltages of the buses in the previous PF calculations. The estimating of the basin of attraction for voltage including the center and, maximum and minimum radius of region are calculated for a system with 7 buses and has been illustrated in Fig. \ref{fig:Basin_of_attraction}. This method is versatile and can be applied to grids of any size and configuration, regardless of the number of buses \cite{Garces-Ruiz2024-qx}. The results can be utilized in combination with other methods, directly as an initialization for NR method (warm start), or even as an independent method.

\subsection{Solution Prediction Pipeline}\label{subsec:learning_approximations}

\begin{figure}[t!]
    \centering
    \includegraphics[width=1\linewidth]{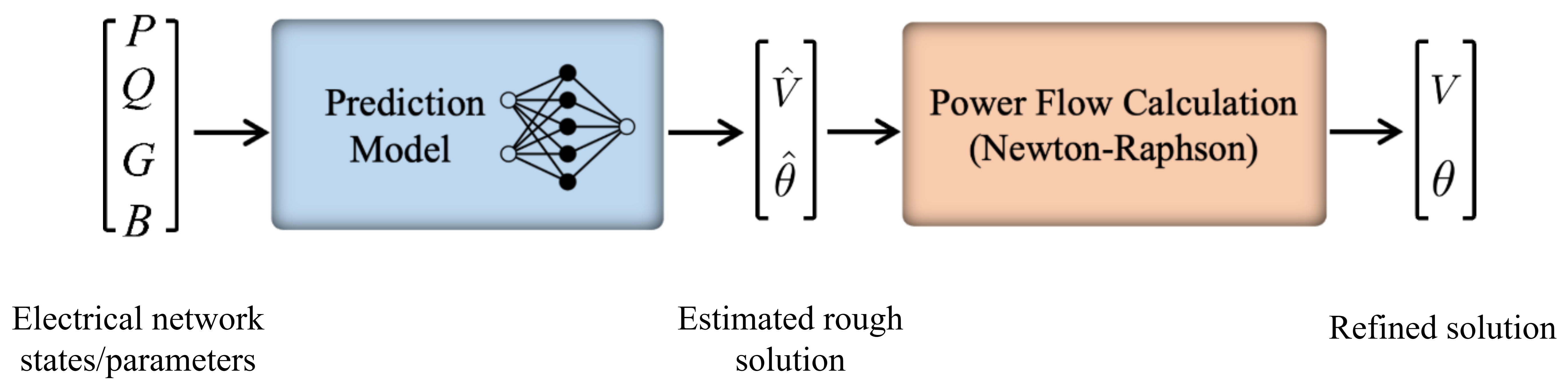}
    \caption{Solution prediction pipeline for obtaining the power flow solution.}
    \label{fig:supervised_pipeline}
\end{figure}

The Alliander case study aims to find an initial solution for power flow estimation. A proper initial solution lies within the basin of attraction, meaning it is as close as possible to the final Newton-Raphson solution. Accordingly, the solution prediction pipeline aims to develop a model for power flow estimation that approximates the NR solution with high precision.  
In this research, neural networks (NNs) are used to model power flow for a two-bus system. The NN-based model generates an estimated solution based on the system state and parameters. In practice, since NNs can only roughly approximate the Newton-Raphson solution, the estimated solution is further refined using the Newton-Raphson method to obtain the final power flow calculation. Fig.~\ref{fig:supervised_pipeline} illustrates the proposed pipeline.  

Two approaches are explored for model development. The first approach utilizes a supervised NN regression model, while the second is an unsupervised method based on PINNs. The details of these approaches will be discussed in the following sections.  

\subsubsection{Regression supervised learning approach}

The regression model approximates the power flow solution and is mathematically expressed as:  
\begin{equation}  
    [\hat{V},\hat{\theta}]=\hat{f}(x), 
\end{equation}  
where the input vector $x = [P, Q, G, B]$ represents the network state and parameters. The model is trained using the mean squared error (MSE) loss function, defined as:  
\begin{equation}  
    \small
    MSE=\frac{1}{N}  \left [ (V-\hat{V})^2 + (\theta-\hat{\theta})^2 \right ].
\end{equation}  

The training dataset consists of 100 two-bus systems with different parameters, each with 10 operational states per configuration. Experimental results demonstrate that the model generalizes well within the two-bus system case and reduces the computational effort required for Newton-Raphson to only 2 iterative steps.  
However, the model shows limitations when applied to larger network topologies.

\subsubsection{Physics-informed unsupervised learning approach}

Another promising approach for estimating a good initial guess is Physics-Informed Neural Networks (PINNs). PINNs \cite{pinn} was first proposed as a data-driven surrogate model for numerical solutions of Partial Differential Equations (PDEs) that injects the physics knowledge represented by PDEs into the learning process of deep neural networks. Inspired by the original work of PINNs, which computes partial derivative of neural network's output components against input variable via an extra backpropagation during training to construct a PDE residual term as the physics-informed loss function, we similarly also construct physics-informed loss function with the input and output of a neural network.
\begin{figure}[t!]
    \centering
    \includegraphics[trim=0.3cm 0 0 0, clip, width=0.8\linewidth]{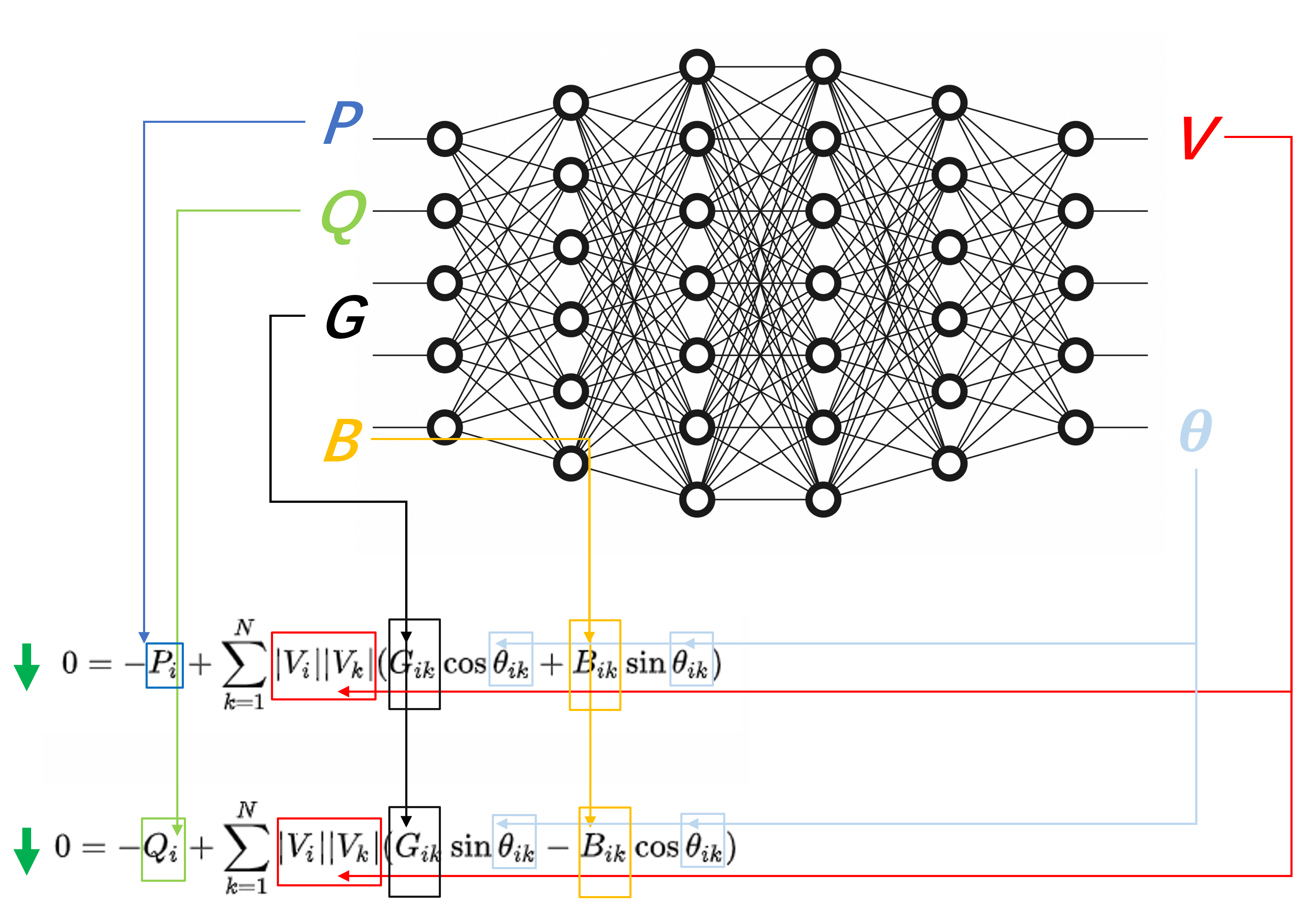}
    \caption{Physics-informed unsupervised learning approach for Newton-Raphson initialization.}
    \label{fig_yan}
\end{figure}

As shown in Fig. \ref{fig_yan}, during training, we take the neural network's input $P$, $Q$, $G$, and $B$ and corresponding output $V$ and $\theta$ and construct the residual terms of Power Flow (PF) Equations, which is to move all terms of PF equations to the right side so that the left side should be zero. The summation of the modulus of these residual terms are used as loss function so that the absolute values of the residual terms are respectively minimized by the optimizer to satisfy the PF equations. Since the first bus is a slack bus, which is arbitrary and should be ignored during power flow optimization, the number of buses $M$ starts from 2 rather than 1. Here, we notice a difference from original work of PINNs that our physics loss does not need an extra backpropagation because no derivative occurs in PF equations.

\begin{small}
\begin{equation}
\begin{split}
    l = \sum_{i=2}^{M} | [ -P_i + \sum_{k=1}^N |V_i||V_k|( G_{(i,k)}\cos \theta_{(i,k)} + B_{(i,k)}\sin \theta_{(i,k)} ) ] | +
    \\ \sum_{i=2}^{M} | [ -Q_i + \sum_{k=1}^N |V_i||V_k|( G_{(i,k)}\sin \theta_{(i,k)} - B_{(i,k)}\cos \theta_{(i,k)} ) ] |
\end{split}
\end{equation}
\end{small}

In practice, grid operators are interested in good initial guesses for the Newton-Raphson to initiate the iterations rather than an accurate solution for PF equations. Therefore, we employ an unsupervised learning scheme, which is also a difference from the original PINNs. During training, we take $P$, $Q$, $G$, and $B$ from the data generation as the input to the neural network, but we don't use the labels $V$ and $\theta$ to construct a data loss to supervise the training. Instead, we just take whatever that is output by the neural network as predicted $V$ and $\theta$. We take input $P$, $Q$, $G$, and $B$ and prediction $V$ and $\theta$ and construct the physics loss. The neural network will only be supervised by physics loss without the labels from the generated data. With this physics-informed unsupervised learning approach, the neural network is learning to predict $V$ and $\theta$ values that only satisfy the PF equations. The assumption here is that the predictions for $V$ and $\theta$ are only rough approximations to the solution labels in the generated dataset, as they have only physical significance but lack alignment with the labels in the generated dataset. Thus, after unsupervised training, the neural network should be able to predict $V$ and $\theta$ values that lie in the attractive basin of a certain $P$, $Q$, $G$, and $B$ as input.

\begin{lemma}
Let $\theta_i \in [0, 2\pi]$ and $V_i \in R$ be the voltage angle and voltage magnitude of bus $i$ satisfying the following equations:
\begin{enumerate}
    \item \begin{flushleft} $0 \approx -P_i + \sum_{k=1}^N |V_i||V_k|( G_{(i,k)}\cos \theta_{(i,k)} + B_{(i,k)}\sin \theta_{(i,k)} )$ \end{flushleft}
    \item \begin{flushleft} $0 \approx -Q_i + \sum_{k=1}^N |V_i||V_k|( G_{(i,k)}\sin \theta_{(i,k)} - B_{(i,k)}\cos \theta_{(i,k)} )$ \end{flushleft}
\end{enumerate}
Then, for $i=2,3,...,M$ all together, the $M \times N$ matrices $\theta_{i,k}$ and $V_{i,k}$ as the initial guess for Newton-Raphson method has a finite number of iterations to converge.
\end{lemma}

After training, we subsequently input the $P$, $Q$, $G$, and $B$ from 100 test samples and use the predicted $V$ and $\theta$ values as the initial guesses for Newton-Raphson method to optimize for these 100 test samples. We compare the iterations for convergence of Newton-Raphson and evaluation metrics averaged over 100 test samples in terms of the unsupervised learning scheme described above, semisupervised learning scheme that uses both data loss and physics loss, and supervised learning that only uses data loss in Tab. \ref{table_yan}. MAE refers to the mean absolute error (MAE) between neural network's predicted and ground truth voltage magnitude or predicted and ground truth voltage angle. Low MAE demonstrates that the output of trained neural network is close to ground truth and vice versa. What is worth mentioning is that when we employ the commonly used zero initialization strategy which assigns zeros for all values in $V$ and $\theta$ as the initial guesses for Newton-Raphson method, Newton-Raphson method failed to converge for all 100 test samples, i.e., cannot converge after 50 iterations.
\begin{table}[t!]
    \centering
    \caption{Comparison of Learning Supervision Schemes}
    \label{table_yan}
    \begin{tabular}{p{1.55cm}|p{1.55cm}|p{1.55cm}|p{1.55cm}}
        \hline
        Metrics & Supervised learning & Semisupervised learning & Unsupervised learning \\
        \hline
        Iterations & 2.73 & 2.80 & 6.88 \\
        \hline
        MAE ($V_{\text{pred}}, V_{\text{NR}}$) & 0.0492 & 0.0526 & 0.5034 \\
        \hline
        MAE ($\theta_{\text{pred}}, \theta_{\text{NR}}$) & 0.0207 & 0.0175 & 29.83 \\
        \hline
        PF equation residual & $8.33e-09+4.74e-07j$ & $1.54e-08+5.98e-07j$ & $5.99e-07+4.42e-07j$ \\
        \hline
    \end{tabular}
\end{table}

\begin{figure}[t!]
    \centering
    \includegraphics[width=0.8\linewidth]{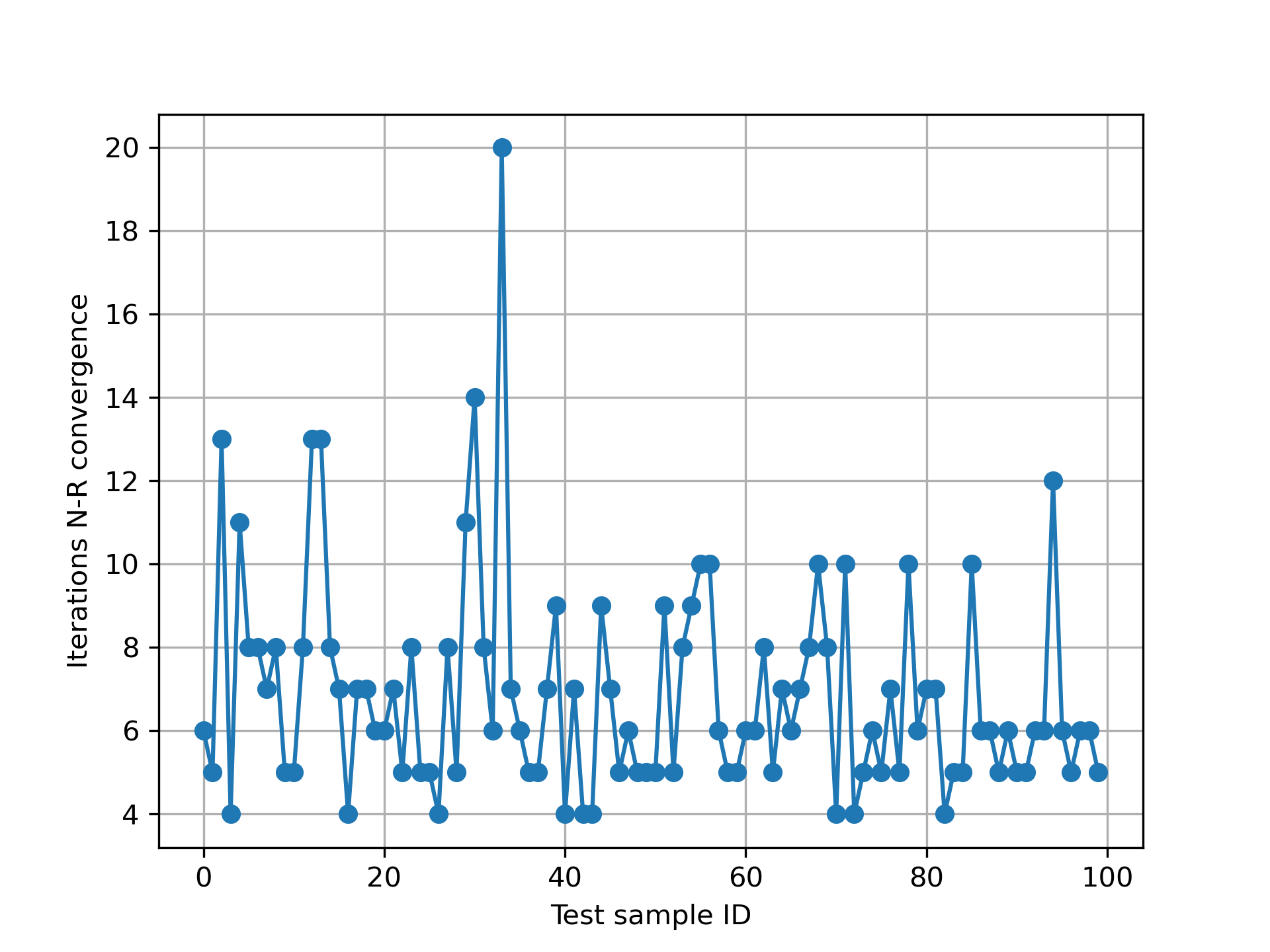}
    \caption{Number of iterations for Newton-Raphson method to converge on 100 test samples with the initial guesses provided by physics-informed unsupervised learning scheme.}
    \label{across_testset}
\end{figure}

The number of iterations needed by Newton-Raphson method to converge on every sample in the test set provided with the initial guess for $V$ and $\theta$ values as the output of the unsupervised learning scheme are shown in Fig. \ref{across_testset}. From Fig. \ref{across_testset}, it is clear that all 100 test samples converged successfully within 20 iterations. Most iterations concentrated between 4 and 10, which are acceptable values in Alliander's industrial practice. The maximum iteration across test set is 20. As a worst-case scenario, 20 iterations is also manageable in a reasonable time. To sum up the PINN approach, the unsupervised learning approach works well for the provided case.

\subsection{Reinforcement Learning}\label{subsec:RL_based_updates}
Reinforcement Learning (RL) is a subfield of machine learning in which an agent learns an optimal decision-making strategy, known as a policy, through interactions with an environment. The agent takes actions based on observations and receives feedback in the form of rewards, which guide the learning process. 
A typical RL framework is formulated as a Markov Decision Process (MDP), which is defined by the tuple: 
\begin{equation}
    \mathcal{M} = \langle S, A, R, P, \gamma, d_0 \rangle,
\end{equation}
where $S$ is the set of states representing the environment's possible configurations, $A$ is the set of available actions, $R: S \times A \times S \to \mathbb{R}$ is the reward function that provides the feedback to the agent, $P: S \times A \to S$ describes the transition dynamics of the environment, $\gamma \in (0,1)$ is the discount factor that balances immediate and future rewards, and $d_0 \in S$ represents the initial state distribution. 
The objective of an RL agent is to learn a policy $\pi: S \to A, s \mapsto \pi(a|s)$ that maximizes the expected cumulative discounted reward over time:
\begin{equation} \label{eq: RL_objective}
    J(\pi) = \mathbb{E}_{\tau \sim \pi} \left[ \sum_{t=0}^{\infty} \gamma^t R(s_t, a_t) \right],
\end{equation}
where $\tau = {s_0, a_0, \dots, s_t, a_t, \dots}$ denotes a trajectory generated by following the policy $\pi$ \cite{sutton2018reinforcement}
. The optimal policy is then defined as:
\begin{equation}
    \pi^{*}=\arg\max_{\pi}J(\pi).
\end{equation}

\subsubsection{2-Bus Case Study}
In our case study, we consider a 2-bus power system, which is characterized by the following MDP parameters:
\begin{itemize}
    \item The initial state distribution $d_0$ consists of an initial guess for $V$ and $\theta$ at the second bus, denoted as $V_0 \in (0.5,2]$ and $\theta_0 \in (-90,90]$ degrees. All initial constant values further characterizing the initial state distribution are shown in table \ref{tab:system_params}.

    \begin{table}[t!]
    \centering
    \caption{System parameters for the experiment.}
        \label{tab:system_params}
        \begin{tabular}{l c}
            \hline
            \textbf{Parameter} & \textbf{Value} \\
            \hline
            External grid voltage ($V_{\textrm{ext}}$) & $1.0$ \\
            $G$ & $100$ \\
            $B$ & $10$ \\
            Active power ($P$) & $0.9$ \\
            Reactive power ($Q$) & $0.6$ \\
            Bus 1 voltage magnitude ($V_{\textrm{bus}_1}$) & $1.0$ \\
            Bus 1 voltage angle ($\theta_{\textrm{bus}_1}$) & $0$ \\
            \hline
        \end{tabular}
    \end{table}
    \item The state $s_t$ at timestep $t \in \{0,...,T\}$ is represented as a vector $[V_t, \theta_t, k_t]$ containing the current voltage magnitude $V_t \in (0.5,2]$, the current voltage angle $\theta_t \in (-90,90]$ degrees, and the number of NR iterations $k_t \in [0,50]$ required to converge.
    \item The actions consist of a vector $a_t = [\Delta V_t, \Delta \theta_t]$, where $\Delta V_t \in (-0.5,0.5]$ and $\Delta \theta_t \in (-50,50]$ degrees represent the voltage magnitude and angle adjustments at timestep $t$, respectively. 
    \item The reward $r_t$ is defined solely by a penalizing term based on the norm of the residual vector $[\Delta P_t, \Delta Q_t]$, where $\Delta P_t$ and $\Delta Q_t$ are the active and reactive power mismatches resulting from applying the updated guesses in the power flow calculations:
    \begin{equation}
        r_t = -||[\Delta P_t,\Delta Q_t]||,
    \end{equation}
    where $r_t$ refers to $R(s_t, a_t)$ in Eq. \ref{eq: RL_objective}.
\end{itemize}
For this case study we let the agent learn a policy that aims to move initial guesses towards values for $V$ and $\theta$ from which it would take NR $k=3$ iterations to converge. Furthermore, for each training episode, the agent would get a maximum of $T=10$ timesteps to take actions before the environment gets terminated.
To optimize the agent's policy, we employ \texttt{stable-baselines3}'s Proximal Policy Optimization (PPO) \cite{Schulman2017PPO} \cite{stable-baselines3} and train for $2\cdot 10^6$ environment timesteps. All hyperparameters used for training the agent can be found in table \ref{tab:ppo_hyperparams}.

\begin{table}[t!]
    \centering
    \caption{Hyperparameter settings for training the PPO agent.}
    \label{tab:ppo_hyperparams}
    \begin{tabular}{l c}
        \hline
        \textbf{Hyperparameter} & \textbf{Value} \\
        \hline
        Learning rate ($\alpha$) & $10^{-4}$ \\
        Discount factor ($\gamma$) & $0.99$ \\
        Number of steps & $2048$ \\
        Batch size & $64$ \\
        Number of epochs & $10$ \\
        Clipping range & $0.2$ \\
        Entropy coefficient & $0.0$ \\
        \hline
    \end{tabular}
\end{table}

\subsubsection{Results}
\begin{figure}[t!]
    \centering
    \includegraphics[width=1.1\linewidth]{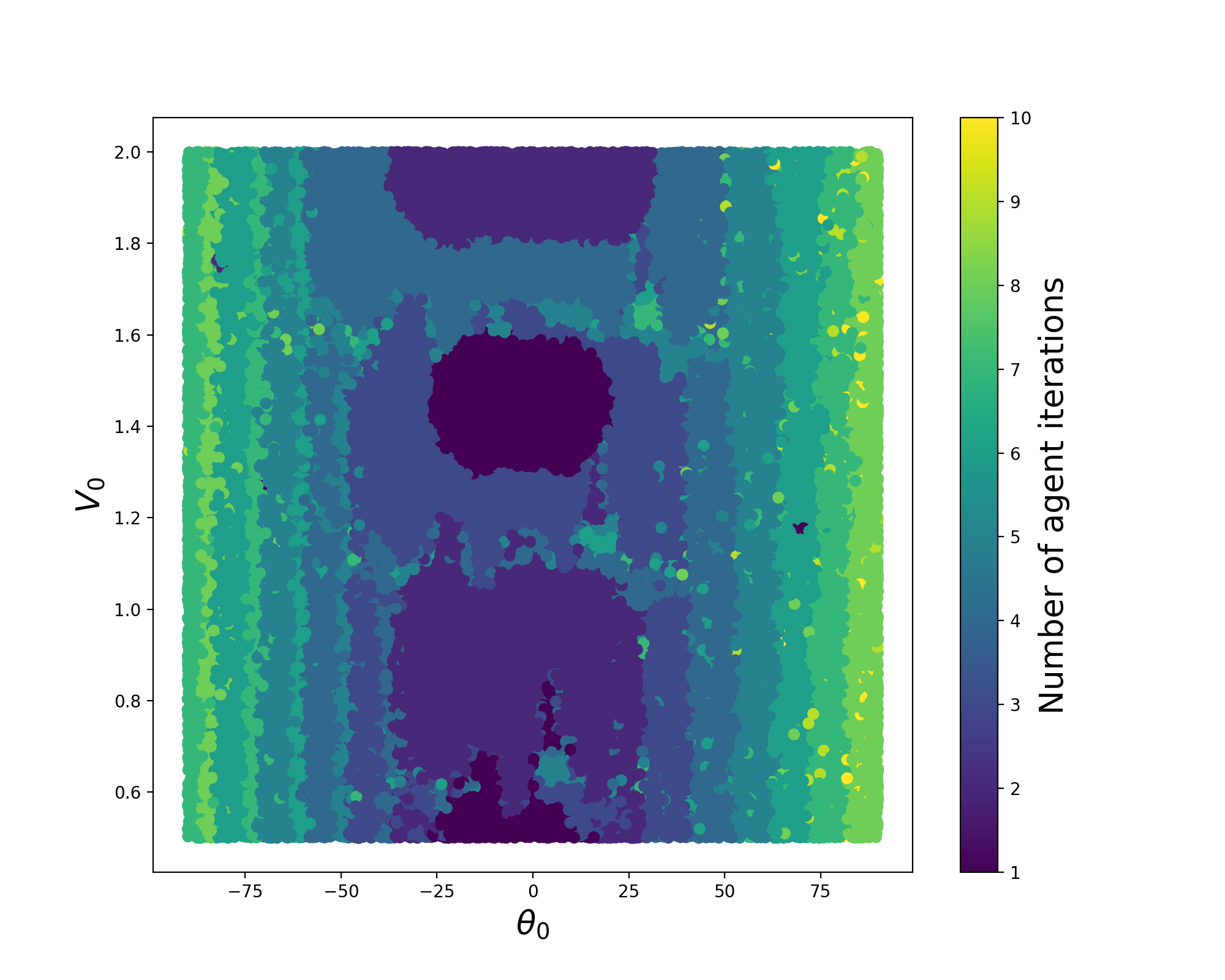}
    \caption{Number of iterations the agent would take to move the initial guesses towards a values from which Newton-Raphson would take $k=3$ iterations to converge.}
    \label{fig:results_timesteps_RL_agent}
\end{figure}

\begin{figure}[t!]
    \centering
    \includegraphics[width=1.1\linewidth]{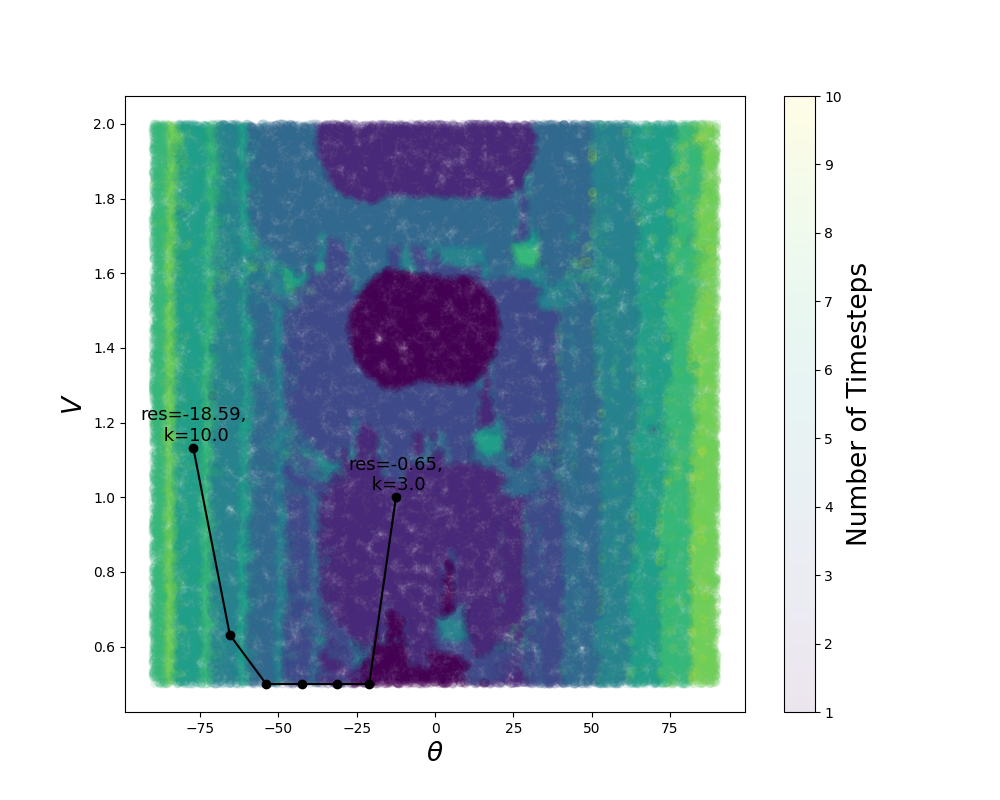}
    \caption{Example of a trained RL agent's executed policy. The agent gets represented with initial values for $V$ and $\theta$ that would take Newton-Raphson $k=10$ iterations, and within 6 timesteps the agent shifts the values for $V$ and $\theta$ towards a region where Newton-Raphson would take $k=3$ iterations to converge.}
    \label{fig:results_trace_RL_agent}
\end{figure}

Fig. \ref{fig:results_timesteps_RL_agent} shows for all $V_0$ and $\theta_0$ drawn from the initial state distribution, the number of timesteps the trained agent would take to update $V$ and $\theta$ such that it would take NR $k=3$ iterations to converge. A maximum of 10 iterations represents the maximum episode length $T=10$. Furthermore, Fig. \ref{fig:results_trace_RL_agent} shows for the trained policy an execution of its strategy. Both the residual and the number of NR iterations get reduced within the maximum episode length. These results indicate that for a 2-bus case, the agent manages to find a policy that could function as an aiding component for PF calculations to converge. 

\section{Future Work}\label{sec:Futurework}
One challenge in the analytical method is that we have a contraction constant that directly affects the radius of the basin of attraction. Defining a method to select an optimum value for this parameter can help achieve optimal boundaries and even the basin of attraction with optimum results.

Future research for regression supervised learning approach could explore replacing the current multilayer perceptron (MLP) model with Graph Neural Networks (GNNs) to better handle high-dimensional data and improve scalability. Given that power grids naturally form graphs, GNNs excel at capturing spatial dependencies by aggregating information from neighboring buses and transmission lines. This enables a more effective modeling of energy flow and system interactions. Additionally, GNNs can adapt to changes in grid topology and operating conditions, making them well-suited for dynamic power systems.

In terms of physics-informed unsupervised learning approach, it is currently successful with 2-bus system. However, in future, we need to train and test on a 7-bus system, or even a system with much larger scale to make it useful in reality. The advantage of this method is that it does not need labeled data, which is time-efficient and cost-efficient in industrial practices. It is not too difficult for Alliander to collect a large amount of $P$, $Q$, $G$, $B$ data, but it is very difficult to compute $V$ and $\theta$ labels using NR optimization iterations for a large amount of data. The disadvantage of this method is that its computational complexity increases exponentially as the input dimension increases, meaning when the electricity grid grows larger and more complex, the training and inference of this method will become more and more difficult and slow. If Alliander invests large GPU computation resources for this method, this disadvantage can be relieved.

\section{Conclusion}\label{sec:conclusion}

This study introduces three approaches to obtain a proper initialization for Newton-Raphson method in power flow calculations: an analytical method that estimates the basin of attraction, a data-driven model leveraging supervised learning and PINNs to predict optimal initialization, and an RL approach to accelerate convergence by incrementally adjusting voltages. 

The analytical method was tested on a 7-bus system, where the minimum and maximum radius of the solution voltage were estimated accurately, reducing the total number of required iterations compared to a random guess. This method is versatile and fast, and can be applied to grids of any size and configuration, regardless of the number of buses, to estimate the basin of attraction and limit the boundaries of the solution. However, it does not provide the final optimum answer and can be used to choose a random guess inside the basin of attraction for better initialization of the NR method or in combination with other methods. 

NN models are another solution for the optimum initialization of the NR method. The regression supervised learning, specifically a MLP model, was trained with a dataset related to a 2-bus test system. The results show good generalization of the model and can reduce the total required iterations of NR to only two. PINNs are another proposed method as an unsupervised learning approach. In this case, the results of unsupervised, semi-supervised, and supervised learning were compared, showing that the unsupervised learning approach can yield reasonable results for the defined case. 

RL is another approach that can be used to adjust the values, $V$ and $\theta$ in this case, with initial guesses, moving them to a region where the NR method can converge within a desired number of iterations. This method was trained on a 2-bus system. According to the test results, the model was able to shift the values from a region requiring 10 iterations to a region requiring only 3 iterations to converge.

\section*{Data and codes availability}
\noindent The dataset and python scripts for the proposed methods are available in project repository on GitHub: \href{https://github.com/PowerGridModel/ict-workshop-newton-raphson}{\textcolor{blue}{https://github.com/PowerGridModel/ict-workshop-newton-raphson}}

\section*{Acknowledgment}
This work is part of the ICT with Industry workshop organized by the Lorentz Center and NWO of the Netherlands, which took place from 20 to 24 January 2025 in Leiden. Author Farzad Vazinram's research in this project has received funding from the European Union’s Important Project of Common European Interest – Next Generation Cloud Infrastructure and Services (IPCEI CIS), in collaboration with the Netherlands Enterprise Agency (RVO). Author Zeynab Kaseb's research about reinforcement learning in this paper has received funding from the NWO DATALESS Project (Grant number: 482.20.602). Author Shengyuan Yan has participated in this research and proposed the PINNs unsupervised learning approach in this paper as part of the doctoral training plan in the \href{https://inshape-horizoneurope.eu/}{\textcolor{blue}{EU InShaPe project}} (EU
Funding Nr.: 101058523 — InShaPe), funded by HORIZON EUROPE Framework Programme of European Union.

\bibliographystyle{IEEEtran}
\bibliography{main}
\end{document}